\documentclass[12pt,preprint]{aastex}

\lefthead{S. Pellegrini}
\righthead{Nuclear accretion in nearby galaxies: clues from {\it Chandra}
observations }
 
\def\msun{~M$_{\odot}$}
\catcode`\@=11
\def\gsim{\ifmmode{\mathrel{\mathpalette\@versim>}}
    \else{$\mathrel{\mathpalette\@versim>}$}\fi}
\def\lsim{\ifmmode{\mathrel{\mathpalette\@versim<}}
    \else{$\mathrel{\mathpalette\@versim<}$}\fi}
\def\@versim#1#2{\lower 2.9truept \vbox{\baselineskip 0pt \lineskip
    0.5truept \ialign{$\m@th#1\hfil##\hfil$\crcr#2\crcr\sim\crcr}}}
\catcode`\@=12

\begin{document}

\title{Nuclear Accretion in Galaxies of the Local Universe: Clues from
$Chandra$ Observations }

\author{S. Pellegrini}

\affil{Astronomy Department, Bologna University, Italy;\\
silvia.pellegrini@unibo.it}

\author{ApJ in press; received Nov. 5, 2004, accepted Jan. 28, 2005}

\bigskip

\begin{abstract}

In order to find an explanation for the radiative quiescence of
supermassive black holes in the local Universe, for a sample of nearby
galaxies the most accurate estimates are collected for the mass of a
central black hole ($M_{BH}$), the nuclear X-ray luminosity
$L_{X,nuc}$ and the circumnuclear hot gas density and temperature, by
using $Chandra$ data.  $L_{X,nuc}$ varies by $\sim 3$ orders of
magnitude and does not show a relationship with $M_{BH}$ or with the
Bondi mass accretion rate $\dot M_B$.  $L_{X,nuc}$ is always much
lower than expected if $\dot M_B$ ends in a standard accretion disc
with high radiative efficiency (this instead can be the case of the
active nucleus of Cen A). Radiatively inefficient accretion as in the
standard ADAF modeling may explain the low luminosities of a few
cases; for others, the predicted luminosity is still too high and, in
terms of Eddington-scaled quantities, it is increasingly higher than
observed, for increasing $\dot M_B$.  Variants of the simple
radiatively inefficient scenario including outflow and convection may
reproduce the low emission levels observed, since the amount of matter
actually accreted is reduced considerably. However, the most
promising scenario includes feedback from accretion on the surrounding
gas: this has the important advantages of naturally explaining the
observed lack of relationship between $L_{X,nuc}$, $M_{BH}$ and $\dot
M_B$, and of evading the problem of the fate of the material
accumulating in the central galactic regions over cosmological times.

\end{abstract}


\section{Introduction}

Improved ground based instrumentation and especially the use of $HST$
have shown a widespread presence of central dark objects of $10^7 -
10^9$\msun, most likely supermassive black holes (SMBHs), at the
center of spheroids (bulges and early-type galaxies) in the local
Universe (Magorrian et al. 1998, van der Marel 1999, Gebhardt et
al. 2003). A possible relationship between these galaxies and the
relics of the ``quasar era'' has therefore been suggested (Richstone
et al. 1998, Yu and Tremaine 2002).  Most nearby nuclei, though, are
radiatively quiescent or exhibit low levels of activity. For example,
in terms of the Eddington luminosity, while $L/L_{Edd}\sim 1$ in
powerful AGNs, $L/L_{Edd}\sim 10^{-9}$ in SgrA$^*$ (Yuan et al. 2003) 
that hosts a securely measured SMBH mass, and $L/L_{Edd}<10^{-8}$
in the nearby elliptical galaxies NGC1399, 4636 and 4472 (Loewenstein et
al. 2001). In the statistically complete spectroscopic survey of
galaxies with $B_T<12.5$ mag (Ho et al. 1997) only $\sim 40$\% of the
nuclei show line emission that could be explained by accretion. 
This radiative quiescence represents one of the most intriguing
aspects of SMBHs in the local Universe (as already recognized by
Fabian and Canizares 1988).  At the same time, correlations have been
discovered involving the SMBH masses ($M_{BH}$) and global properties
of their host galaxies, as the central stellar velocity dispersion
$\sigma$ (Gebhardt et al. 2000, Ferrarese and Merritt 2000). These
observational facts have led to think that the birth, growth and
activity cycle of SMBHs and the evolution of their host galaxies are
tightly linked. However, it is still under study how the tight
correlations were established, whether the radiative quiescence is
linked to the mechanism responsible for the $M_{BH}-\sigma$ relation,
and why and how the luminous AGNs switched off (e.g., Haiman et
al. 2003).

In this work the question of why local SMBHs are not bright is
addressed. The answer is not in a different value for $M_{BH}$, since
the respective SMBH masses of distant AGNs and local nuclei cover
roughly the same range (Ho 2002). It could then reside in a low mass
accretion rate $\dot M$, or in a low radiative efficiency, or in the
existence of activity cycles. The most promising solution among these
is here looked for by collecting three quantities that play a
fundamental role in this problem (the nuclear emission, $M_{BH}$ and
$\dot M$) for a sample of galactic nuclei in the local Universe, and
by searching for possible relationships among them.  This is
accomplished by using $Chandra$ results for the galactic nuclei and
the best estimates available of their $M_{BH}$.  With the $Chandra$
$\sim 0^{\prime\prime}\hskip -0.1truecm .3$~FWHM PSF (Van Speybroeck
et al. 1997) it has become possible to get a clean look at the faint
nuclear emission that may be associated with SMBHs and also at the hot
gas properties close to the accretion radius, where the dynamics of
the gas start to be dominated by the potential of the SMBH. This
allows us to estimate $\dot M$ in the simplest case of the steady and
spherically symmetrical Bondi (1952) solution ($\dot M_B$). A mass
accretion rate of the order of $\dot M_B$ enters also {in the
viscous rotating analog of the Bondi treatment represented by radiatively
inefficient accretion flow models (Narayan and Yi 1995, Quataert 2003)}.

In sect. 2 the collected sample is presented, spanning
morphological types from E to Sbc. In Sect. 3 the mass accretion rate
$\dot M_B$ is derived in a homogeneous way for a subsample of
early-type galaxies, and the possible sources of uncertainty are
discussed; then, the relationship between $L_{X,nuc}$, $M_{BH}$
and $\dot M_B$ are investigated. In Sect. 4 the observational findings
are summarized and compared with the predictions of various models for
low luminosity accretion. In Sect. 5 the results are discussed further.

\section{The sample}

For this study all nearby nuclei with a Chandra investigation of
their nuclear luminosity $L_{X,nuc}$ are considered. A few nuclei
classified as Seyfert (Ho et al. 1997), or residing in peculiar
objects as starburst and closely interacting systems, or with a highly
uncertain $L_{X,nuc}$ estimate (e.g., due to severe pile-up problems)
have been excluded. The elliptical Cen A, the nearest active galaxy,
is added to the sample for comparison and later reference. In order
for a nucleus to be included, its $M_{BH}$ estimate must derive from
specific modeling (e.g., Gebhardt et al. 2003 for NGC4697) or it must be
possible to calculate it from the $M_{BH}-\sigma$ relation of Tremaine
et al. (2002), for the proper $\sigma $ value [this is derived from
McElroy (1995) or the HyperLeda catalogue]. The $M_{BH}-\sigma$
relation has an intrinsic dispersion in $M_{BH}$ within a factor of
two (Tremaine et al. 2002).  The resulting 50 host
galaxies are listed in Table 1; their morphological types go from E0
to Sbc (as shown by Fig. 1 discussed in Sect. 3.1).  The circumnuclear
hot gas density $\rho$ and temperature $T$ have been derived from a
$Chandra$ pointing for 17 of the galaxies in Table 1, in addition to
the Galactic Center. All these are early-type systems: 15 are E or S0
galaxies, 2 of them are Sa (Sombrero and NGC1291).

Since distance-dependent quantities are involved in this work,
distances obtained in a homogeneous way have been adopted (column 2),
as are those derived from the SBF method by Tonry et al. (2001); 
this is possible for most of the galaxies, for the others (labeled in
Tab. 1) the adopted distance refers to $H_0=75$ km s$^{-1}$
Mpc$^{-1}$, a value consistent with the $H_0$ implied by Tonry et
al. (2001). The values of $T$, $\rho$ and $L_{X,nuc}$ in columns 5,
6, and 7 have been derived by the authors referenced in column 8, and
have been rescaled for the distance in column 2 when necessary. The
values of $M_{BH}$ in column 3 have been rescaled for the distance in
column 2 when needed; the source of the $M_{BH}$ estimate is given in
column 4.

$L_{X,nuc}$ is that of a point source located at the optical or radio
center of the galaxy. In most cases, the nuclear emission is hard and
its spectral distribution can be modeled with a power law of photon
index $\Gamma=1-2$. The uncertainty on $L_{X,nuc}$ is typically well
within 20\%; in 11 cases just an upper limit could be placed on
$L_{X,nuc}$. The values of $\rho$ and $T$ in Tab. 1 refer to the
accretion radius $r_{acc}=2GM_{BH}/c_s(\infty)^2$, where $c_s(\infty)$
is a ``fiducial'' sound speed of the ISM [i.e., valid in the
circumnuclear region; see eq. (2.38) of Frank et al. 2002]. At
$r_{acc}$ the ratio of internal energy to gravitational binding energy
of a gas element is $\sim 1$ and therefore for $r\lsim r_{acc}$ the
gravitational pull of the SMBH is the prevailing force on the
surrounding ISM (if heating sources can be neglected). For the
selection of nuclei with $\rho$ and $T$, a distance limit of 50 Mpc
has been adopted, in order for the $Chandra$ ACIS angular resolution
to provide a reasonable measurement (a typical $r_{acc}$ of $ 100$ pc
corresponds to 0.4 arcsec at 50 Mpc). In this way the $\rho$ and $T$
values in Tab. 1 have been directly estimated at or extrapolated
reasonably well to their $r_{acc}$ (whose angular size ranges from
$0^{\prime\prime}\hskip -0.1truecm .2$ to $2^{\prime\prime}$), by
deprojection of X-ray imaging and spectroscopic data. For four
galaxies in Tab. 1, instead, they are an average over a central region
whose radius is much larger than $r_{acc}$ (M32, NGC821, NGC1553 and
NGC4438)\footnote{ For example, for M32 they refer to an annulus of
radii of 15$^{\prime\prime}$ and 44$^{\prime\prime}$, they are an
average over a central region of projected radius of
$20^{\prime\prime}$ for NGC821, and an average over a central
spiral-like region extending $\sim 30^{\prime\prime}$ for NGC1553 (see
the references in column 8).}. How are these $\rho$ and $T$ likely to
vary if calculated at $r_{acc}$?  In the cases studied best with
$Chandra$, $T$ does not vary more than $ 50$\% in the central galactic
region, being usually decreasing towards the center (see, e.g., Di
Matteo et al. 2003 for M87, Kim and Fabbiano 2003 for NGC1316, Ohto et
al. 2003 for NGC4636); the radial density distribution, instead,
always raises smoothly down to the smallest observed radii, with an
increase by a factor of a few or more. The effect of this uncertainty
for these four galaxies is taken into account when $\rho$ and $T$ are
used below (Sects. 3.2--3.4).

\section{Results}

\subsection{$L_{X,nuc}$ and $M_{BH}$}

The relationship between the nuclear X-ray luminosity and the SMBH
mass is shown in Fig. 1.  No clear trend between these two quantities
is apparent from this plot. A lack of nuclei with high
$L_{X,nuc}$ and $M_{BH}$ between $1$ and $5\times 10^7$\msun$\,$ can be
seen, but its real existence should be further checked 
with larger samples when available. On the contrary it is clear that, for
$M_{BH}>5\times 10^7$\msun, $L_{X,nuc}$ varies by a large factor,
roughly three orders of magnitude, at any fixed $M_{BH}$. 

The nucleus with the highest $L_{X,nuc}$ in Fig. 1 is that of the
active galaxy Cen A (see Tab. 1). Excluded Cen A, the six brightest
nuclei are those of NGC3169, NGC3226, NGC4261, NGC4486, IC1459,
IC4296. From their optical emission line spectra, these are classified
as LINERs or show weak or absent optical lines (Ho et al. 1997,
Phillips et al. 1986, Wills et al. 2002).  The last four of these are
also radio galaxies, the first two are not. Note that there are other
radio galaxies in Fig. 1 (e.g., NGC1316 and NGC4374) that instead have
$L_{X,nuc}<10^{40}$ erg s$^{-1}$.  In this respect it is interesting
to note that the core radio luminosity of nearby galactic nuclei shows
a similar large variation and lack of relation with the SMBH mass for
$M_{BH}\sim 10^7-10^9$\msun$\,$ (Ho 2002).

\subsection{$L_{X,nuc}$ and $\dot M_{B}$}

The next interesting relationship to investigate is that between the
nuclear luminosity and the mass potentially available for
accretion. This makes use of $\rho$ and $T$ collected in Table 1.  The
simplest assumption to make is that gas accretion is steady and
spherically symmetric as in the standard theory developed for gas
accreting onto a point mass at rest with respect to it (Bondi
1952). In this theory the accretion rate $\dot M_B$ is given
by [see eq. (2.36) of Frank et al. 2002]

\begin{equation}
\dot M_B=\pi G^2 M_{BH}^2 {\rho (\infty)\over c_s^3(\infty)}\left[
{2\over 5-3\gamma}\right]^{(5-3\gamma)/2(\gamma-1)} 
\end{equation}
where $\gamma$ is the polytropic index that
varies from 1, in the isothermal case, to 5/3 in the adiabatic case;
$c_s=\sqrt{\gamma kT/\mu m_p}$ is the sound speed of the gas, with
$m_p$ the proton mass and $\mu$ the mean mass per particle of gas
measured in units of $m_p$; $\mu $ is assumed here to be equal to
0.62, corresponding to a solar chemical composition; finally,
``$\infty $'' refers to the ambient conditions.  It is usually assumed 
that the accretion rate is determined by $\rho$ and $T$ at the radius
where the influence of the black hole becomes dominant (i.e., close to
$r_{acc}$ defined in Sect. 2).  The relationship between $L_{X,nuc}$
and $\dot M_B$ is shown in Fig. 2, for $\gamma=1.33$ (an intermediate
value between the two limits). For the cases where $\rho$ is likely an
underestimate of $\rho (r_{acc})$ and $T$ could overestimate
$T(r_{acc})$ (as discussed at the end of Sect. 2), $\dot M_B$ derived
here is likely an underestimate of the true $\dot M_B$ [from eq. (1)]
and as such is marked in Fig. 2. No clear trend between
$L_{X,nuc}$ and $\dot M_B$ is shown by Fig. 2; a scatter of $\sim 3$
orders of magnitude is shown by both $L_{X,nuc}$ and $\dot M_B$.

For a few of the nuclei in Fig. 2 a ``Bondi mass accretion rate'' had
already been calculated by the authors referenced in Table 1 (col. 8),
by using though slightly different definitions for $\dot M_B$,
different ways of estimating $M_{BH}$ and the gas density $\rho$ to be
inserted in eq. (1) from observational data, different values for
$\gamma$ and $\mu $, and distances derived with different methods and
not referring to the same distance scale.
A re-calculation of $\dot M_B$ in a homogeneous way was needed here in
order to consider the problem of the nature of accretion for these
nuclei as a class.  Unfortunately, residual uncertainties remain on
the $\dot M_B$ derived here. From eq. (1), they are due to the
errors on the estimate of $\rho(r_{acc})$, $T(r_{acc})$ and $M_{BH}$;
in addition, the possible range of values for $\gamma$ between 1 and
5/3 causes $\dot M_B$ to vary by a factor of $9.68$.  The size of the
uncertainty on $\dot M_B$ can be estimated accurately for 7 galaxies
for which errors on $\rho$ and $T$ in Tab. 1 are given in the
literature, by using the standard error propagation formula. The
results are shown in Figs. 2 and 3.  Cen A and Sombrero have the
largest uncertainties on $\dot M_B$, due to the error on the
respective $M_{BH}$ values.

\subsection{$L_{X,nuc}$ and $L_{acc}$}

If at very small radius the accreting gas $\dot M_B$ joins a standard
accretion disc (Shakura and Sunyaev 1973), so that the final stages of
accretion are similar to those of bright AGNs, an accretion luminosity
$L_{acc}\sim 0.1 \dot M_B c^2$ is expected.  This is plotted as a
solid line in Fig. 2.  All the nuclei lie well below this
expectation, which is a representation of the underluminosity problem
for nearby galactic nuclei with the presently best available data. On
the contrary, the nucleus of Cen A (recognizable by its
highest $L_{X,nuc}$ value in Fig. 2) could host a standard disc
with an accretion rate close to $\dot M_B$.  Note that the bolometric
luminosity $L_{bol}$ of the nuclei should be used for a comparison
with $L_{acc}$; however, even when considering $L_{bol}$ instead of
$L_{X,nuc}$, the conclusions are likely to remain unchanged.  For
example, the canonical bolometric correction for AGNs is 
$L_{bol}/L_{X}\sim 10$ (Elvis et al. 1994); more specifically
for the nuclei of
NGC4261, NGC4594 and M87 (three galaxies in Table 1)
the ratio $L_{bol}/L_{0.5-10\,{\rm keV}}$ 
is respectively 14, 8 and 17, calculated from their whole spectral
energy distributions (Ho 1999).  These values do not fix the
underluminosity problem.

\subsection{Relationship between Eddington-scaled quantities }

Fig. 3 shows the relationship between the Eddington-scaled quantities
$L_{X,nuc}/L_{Edd}$ and $\dot M_B/\dot M_{Edd}$ (with $\dot
M_{Edd}=L_{Edd}/0.1 c^2$).  The reason for plotting Eddington-scaled
quantities lies in the possibility of a direct comparison with the
predictions of low radiative efficiency accretion flows (ADAF, Narayan
and Yi 1995).  These can develop in the conditions of very low $\dot
m$ (defined as $\dot m=\dot M/\dot M_{Edd}$), precisely when $\dot m<
\alpha ^2 \lsim 0.1$, where $\alpha$ is a viscosity parameter for the
flow. ADAF models predict a rate of mass accretion $\dot M$ comparable
to the Bondi rate, with a more accurate estimate that may be $\dot
M\approx \alpha \dot M_B$ (Quataert 2003). Given the values of the
abscissae for the points in Fig. 3, these nuclei are candidate to host
ADAFs.  In these flows the matter is so hot and tenuous that it is
unable to radiate strongly; most of the gravitational potential energy
is advected by ions inside the event horizon. The ADAF emission is in
the X-ray and in the radio bands\footnote{The radio emission from an
ADAF scales with the SMBH mass and the X-ray luminosity as
$L_{15\,GHz}\sim 10^{36} (M_{BH}/10^7M_{\odot}) (L_{2-10{\rm
keV}}/10^{40}$erg s$^{-1})^{0.14}$ erg s$^{-1}$ (Yi and Boughn
1999).}, and scales as $L_{ADAF}\sim 0.1\dot M c^2 (\dot m/\alpha^2)$
(Narayan and Yi 1995). This emission level is plotted in Fig. 3 as a
dashed line, for the two cases of $\dot M=\dot M_B$ and $\dot M=\alpha
\dot M_B$, with $\alpha =0.1$ (a value typically assumed for galactic
SMBHs, Di Matteo et al.  2003).  Fig. 3 shows that $L_{ADAF}$ is in
fact much lower than $L_{acc}$, so that an ADAF may explain the
emission level for a few nearby galactic nuclei.
However, $L_{ADAF}$ is still too high for the emission level
of many other nuclei. 
The most discrepant cases are those of the galaxies in the lower right
portion of Fig. 3, that are NGC4472 and NGC1399 (see also Loewenstein
et al. 2001), NGC4649 and the Galactic Center (e.g., Baganoff et
al. 2003, Yuan et al. 2003). Note also how the possibility for an ADAF
to reproduce the observed values of $L_{X,nuc}$ decreases with
increasing $\dot M_B/\dot M_{Edd}$, since $L_{ADAF}/L_{Edd}$ increases
steeply as $\dot m^2$.

\section{Summary }

For a sample of galaxies of the local Universe that excludes Seyferts,
starbursts and peculiar objects, the most accurate estimates available
for $M_{BH}$, $L_{X,nuc}$ and circumnuclear $\rho$, $T$ have been
collected, with the aim of studying possible relationships between
$M_{BH}$, $L_{X,nuc}$ and the Bondi mass accretion rate $\dot
M_B$. It is found that:

$\bullet $ $L_{X,nuc}$ does not show a relationship with
$M_{BH}$. It exhibits a large scatter, up to $\sim 3$ orders of
magnitude, for any fixed $M_{BH}>5\times 10^7 $\msun. Note that
the core {\it radio} luminosity of nearby galactic nuclei shows a
similar large variation and lack of relation with the SMBH mass, for
$M_{BH}\sim 10^7-10^9$\msun (Ho 2002).

$\bullet$ $L_{X,nuc}$ does not show a relationship with $\dot M_B$
either, even though the uncertainties in the latter may be large.
$\dot M_B$ also spans a range of $\sim 3$ orders of magnitude.
 The emission level given by $L_{acc}\sim 0.1 \dot
M_B c^2$, describing the expected emission if $\dot M_B$ ends in a
standard accretion disc, is much larger than the observed $L_{X,nuc}$
values.  Therefore in general, as long as $\dot M_B$ is a good
estimate of the true mass accretion rate $\dot M$, the nuclei are not
downsized AGNs, in the sense that their low level of emission cannot
be accounted for just by a $\dot M<<\dot M_{Edd}$.

$\bullet$ radiatively inefficient accretion as in the standard ADAF
modeling may explain the low luminosities of some nuclei; for 
others, the predicted emission level is still higher than observed. In
addition, the ADAF-predicted luminosity is increasingly higher than
the lowest $L_{X,nuc}$ observed, for increasing $\dot M_B/\dot M_{Edd}$.

Radiatively inefficient scenarios include also `advection dominated
inflow/outflow solutions' (ADIOS, Blandford and Begelman 1999) and
`convection dominated accretion flows' (CDAFs, Quataert and Gruzinov
2000, Igumenshchev et al. 2000), where much less than the mass
available at large radii (i.e., of $\dot M_B$) is actually accreted on
the SMBH. ADIOS prevent accretion by removing completely 
some of the inflowing matter via a polar outflow, in CDAFs accretion
is stalled by convecting the material back out to larger radii. The
accretion of rotating gas may also result in a reduced
$\dot M$ relative to the Bondi rate (Proga and Begelman 2003). All
these variants of the simple radiatively inefficient scenario may be
able to reproduce the low emission levels observed, by
reducing considerably the amount of matter that is actually accreted.

However, the observational evidence of the independence of
$L_{X,nuc}$ from $M_{BH}$ and $\dot M_B$ provided by Figs. 1, 2 and 3
is best explained if there is feedback from the SMBH accretion on the
surrounding ISM. Feedback can be provided by radiative or momentum
driven heating of the circumnuclear gas. In this scenario accretion
undergoes activity cycles: while active, the central engine heats the
surrounding ISM, so that accretion is offset; then the ISM starts
cooling again and accretion resumes.  Intermittent accretion was
already suggested and investigated in the context of the evolution of
galactic cooling flows, and proposed heating sources were the nuclear
hard radiation (Ciotti and Ostriker 2001) or the deposition of the
mechanical energy of nuclear outflows (Binney and Tabor 1995, Omma et
al. 2004). Also in the case of ADIOS the predicted wind may have
an impact on the hot gas at large scales , although this aspect has not
been addressed in detail yet.  The intermittent accretion scenario
could then be described by a series of ADIOS with time dependent outer
boundary conditions (e.g., Yuan et al. 2000).

If there is feedback, and accretion becomes intermittent, the estimate
of $\dot M$ given by a steady spherically symmetric theory without
heating sources, such as the Bondi theory, may be misleading. Also, no
clear relationship of $L_{X,nuc}$ with $\dot M_B$ and $M_{BH}$ is
expected. In a plot like Fig. 2 Cen A could be accreting at the
present time with $\dot M\sim \dot M_B$ and with a high radiative
efficiency; the bulk of the nuclei would be accreting with a largely
different $\dot M$, since they are captured in various stages of their
complex evolution. Nearby early-type galaxies for which a search for
nuclear emission has been made using the $Chandra$ data, but no
detection was found, together with a few of the upper limits on
$L_{X,nuc}$, may correspond to truly inactive SMBHs, where accretion
is temporally switched off due to feedback.

\section{Discussion}

Other, more indirect, arguments favoring a role for feedback
are discussed in this last Section.
 
In favor of the existence of activity cycles is some observational
evidence coming from a galactic scale, where $Chandra$ revealed hot
gas disturbances that are reconducted to the effects of recent nuclear
activity.  For example two symmetric arm-like features cross the
center of NGC4636 (Jones et al. 2002), and are accompanied by a
temperature increase with respect to the surrounding hot ISM; they
were related to shock heating of the ISM, caused by a recent nuclear
outburst.  A similar hot filament crosses the nuclear region of NGC821
(Fabbiano et al. 2004) and a nuclear outflow has been detected in
NGC4438 (Machacek et al. 2004). Cavities and surface brightness edges
related to radio activity have been revealed in NGC4374 (Finoguenov
and Jones 2001) and NGC4472 (Biller et al. 2004).

On the other hand the possibility that radiatively inefficient
accretion takes place after the end of the bright QSO phase 
for cosmological times seems problematic.  If accretion at an
average rate (from Fig. 2) of $\dot M\sim 10^{-2}$\msun yr$^{-1}$
steadily accumulates mass at the galactic centers for $\sim 10$ Gyrs,
then SMBH masses of $\sim 10^8$ \msun$\,$ are formed. SMBHs of masses
$M_{BH}\gsim 10^8$\msun$\,$ already come from accretion with high
radiative efficiency during the optically bright QSO phase (Yu and
Tremaine 2002), therefore after the end of this phase SMBH growth by
accretion with low radiative efficiency at the $\dot M_B$ of Fig. 2
cannot have been very important. Solutions as ADIOS and CDAF predict
an effective $\dot M$ much lower than $\dot M_B$; in this case,
however, it is unclear where does the gas that fails to accrete go,
whether it accumulates in the circumnuclear region and for how long
these solutions can prevent mass from accreting. On the contrary,
feedback modulated accretion has the possibility of displacing gas far
from the galactic center and even removing it from the galaxy;
therefore it presents the advantage of accumulating much less mass at the
galactic centers (e.g., Ciotti and Ostriker 2001). The low
radiative efficiency solutions could however represent a temporary
effect, for example taking place within an intermittent accretion
scenario, or considering that the circumnuclear environment may be
modified by different astrophysical processes over timescales
longer than the accretion time near $r_{acc}$.

Finally, it must also be considered that the galactic ISM has a
substantial and continuous mass input from stellar mass losses. A
robust estimate for this source of mass is $\dot M_*\simeq 1.5\times
10^{-11}L_B(L_{\odot})\,t({\rm 15 Gyr})^{-1.3}$\msun yr$^{-1}$ for an
early-type galaxy of present blue luminosity $L_B$ and age $t$ (valid
after an age of $\sim 0.5 $ Gyr; Ciotti et al. 1991); i.e., $\dot
M_*\simeq 1.5$\msun yr$^{-1}$ in a 15 Gyrs old galaxy of
$L_B=10^{11}L_{\odot}$.  In stationary conditions (or
quasi-stationary, since $\dot M_*$ is a decreasing function of time
$t$) this $\dot M_*$, or a fraction of it coming from the inner few
kpc of the galaxy, feeds a cooling flow towards the galactic center,
if no strong heating sources are present (e.g., Sarazin and White
1988; Pellegrini \& Ciotti 1998).  This has two consequences: 1) in
the past the mass accretion rate towards the galactic center was
likely higher than at the present epoch, since $\dot M_*$ was
higher; therefore in the past we may expect $\dot M_B$ values higher
than in Fig. 2, which constrains the duration of accretion phases
to be even shorter; 2) for a low radiative efficiency scenario, where
much less than $\dot M_B$ can be accreted, to accomodate this
continuous flow of mass towards $r_{acc}$ (and the continuous mass
source from within $r_{acc}$ as well) may represent an additional
problem. Again, feedback modulated accretion shows the advantage of
efficiently removing gas from the galactic centers. This aspect of the
intermittent scenario was in fact previously suggested as a solution
for the well known problem of galactic cooling flows of accumulating
too much cold gas at the galactic centers with respect to the
observations, if lasting for many Gyrs.

In conclusion, the observational evidence provided by $Chandra$
and the best estimates currently possible for the SMBH masses of
nearby galactic nuclei show a large dispersion of the $L_{X,nuc}$ and
$\dot M_B$ values, and lack of relationship between $L_{X,nuc}$, $M_{BH}$,
$\dot M_B$, or their Eddington-scaled values, 
all quantities playing a fundamental role in accretion models.  This
seems to pose a challenge to models not providing feedback to the
ISM, such as the Bondi accretion, ADAF and CDAF. Models with feedback 
may explain the observed scatter
and also provide an efficient way of limiting the accumulation of mass
towards the galactic centers predicted over cosmological times.
 
\acknowledgments
I thank G. Bertin for useful comments, L. Ciotti for
discussions and D.W.  Kim for information about his study on
NGC1316. This work has been partially supported by MIUR (co-fin 2004).

\clearpage

\begin{figure}
\plotone{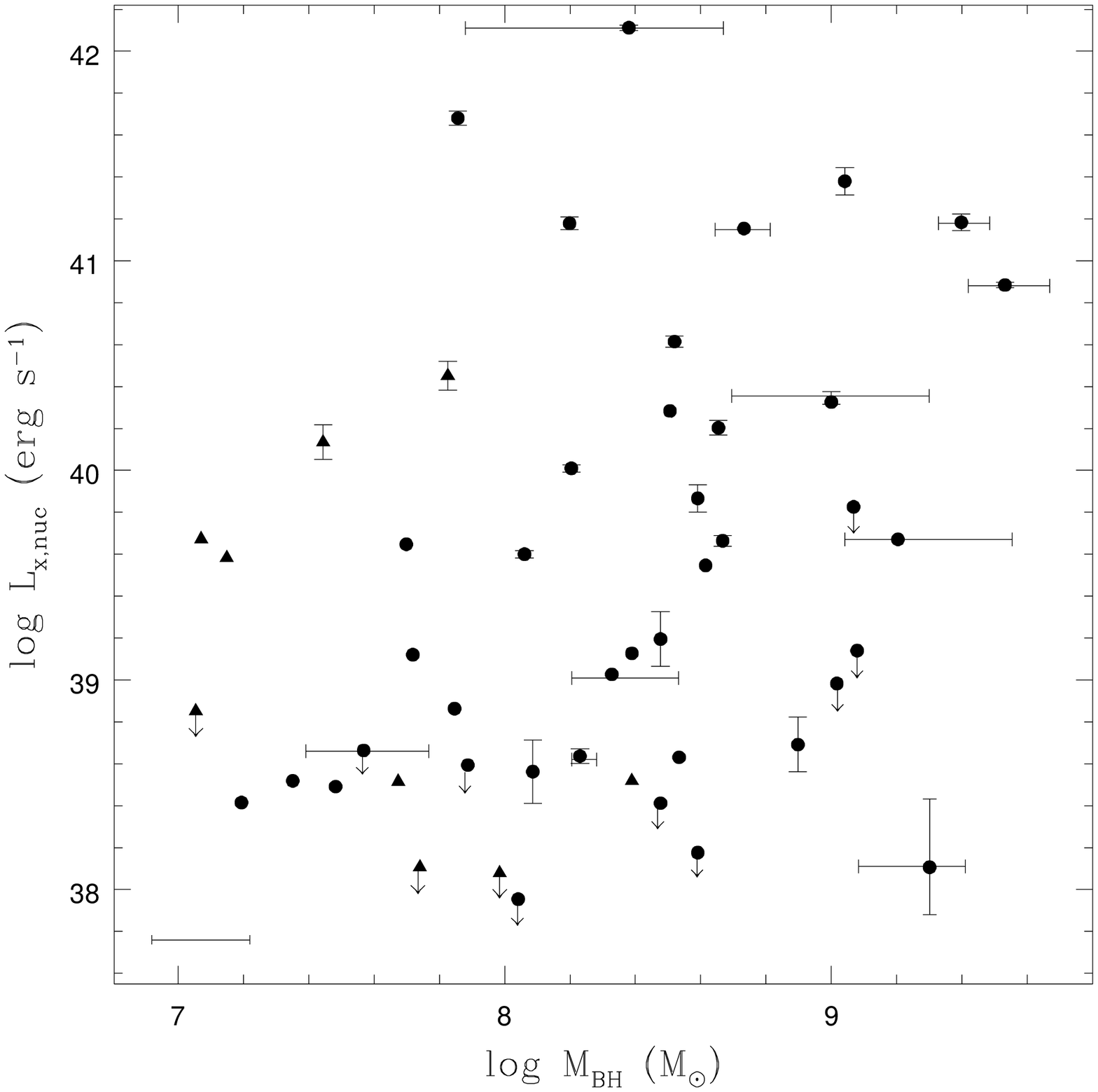}
\hskip -0.6truecm
\caption{The relation between the nuclear X-ray luminosity, as
measured from $Chandra$ data, and the central SMBH mass, for the
galaxies in Table 1 (see Sect. 3.1). Circles indicate morphological
types from E to Sa included, triangles the other types (i.e., from Sab
to Sbc).  Arrows indicate upper limits on $L_{X,nuc}$; uncertainties
on $L_{X,nuc}$ are shown as errorbars when derivable from the
literature.  The uncertainties on $M_{BH}$ are also shown as
errorbars; $M_{BH}$ values estimated from the $M_{BH}-\sigma$ relation
have an uncertainty plotted in the lower left corner (Tremaine et
al. 2002).  M32 and the Galactic Center do not appear in this plot
since their $L_{X,nuc}$ is too low by orders of magnitude (Tab. 1).}
\end{figure}

\clearpage

\begin{figure}
\plotone{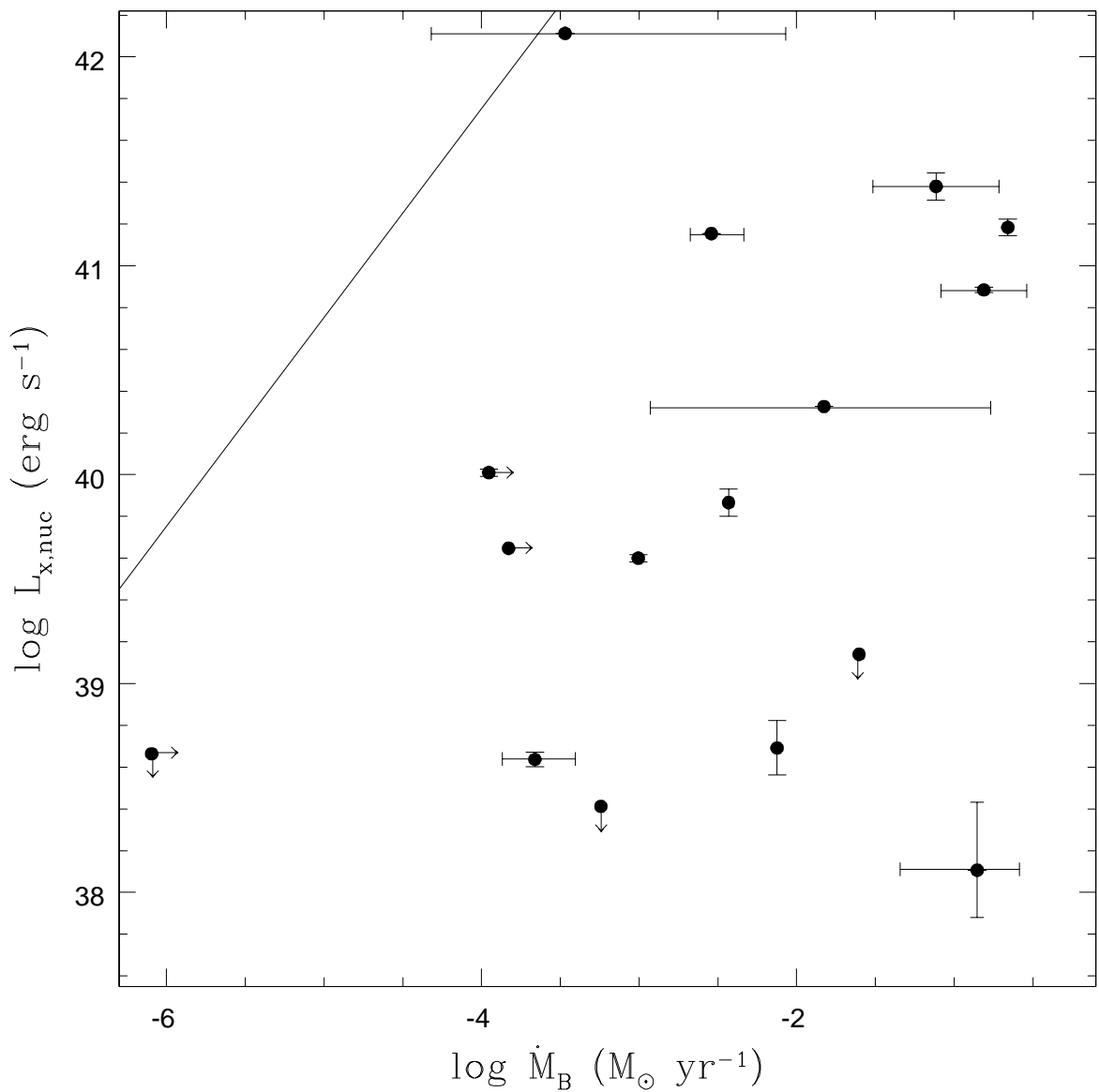}
\caption{The relation between $L_{X,nuc}$ and the Bondi mass accretion
rate $\dot M_B$ estimated as described in Sect. 3.2.  Downward arrows
and errorbars on $L_{X,nuc}$ are as for Fig. 1. Rightward arrows indicate
underestimates of $\dot M_B $ and errorbars its uncertainty,
calculated as described in
Sect. 3.2.  The solid line represents $L_{acc}=0.1 \dot M_B c^2$
(Sect. 3.3). As for Fig. 1, M32 and the Galactic Center cannot appear 
in the plot.}
\end{figure}

\clearpage

\begin{figure}
\plotone{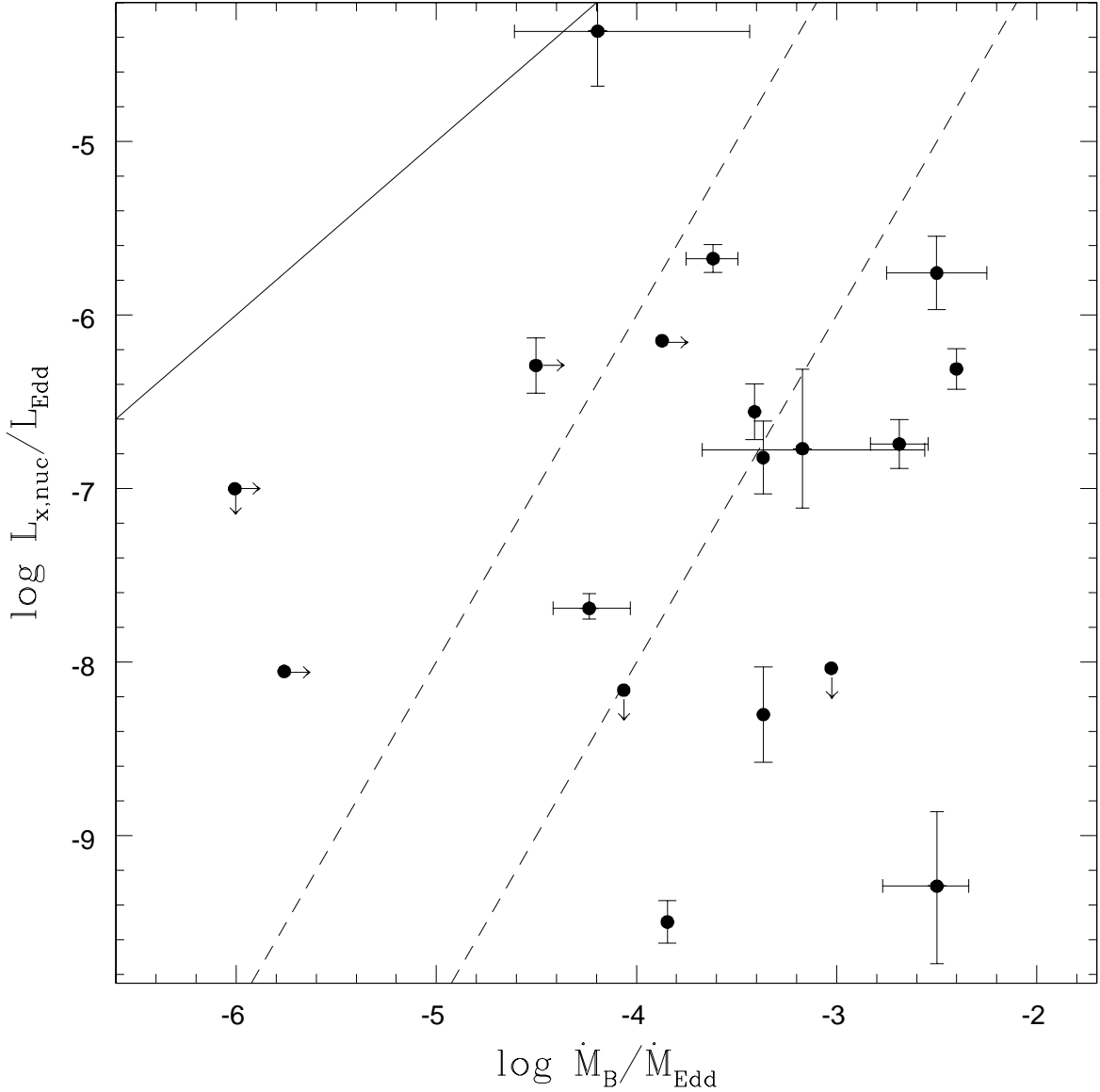}
\caption{The relation between $L_{X,nuc}$ scaled by the Eddington
luminosity and $\dot M_B$ scaled by the Eddington mass accretion rate.
The solid line indicates $L_{acc}/L_{Edd}$; the dashed lines indicate
$L_{ADAF}/L_{Edd}$, the expected emission level for a standard ADAF
model with $\dot M=\dot M_B$ or $\dot M=\alpha \dot M_B$ and
$\alpha=0.1$ (see Sect. 3.4).  The other symbols are the same as
for Fig. 2. The Galactic Center is the point with the lowest $L_{X,nuc}/
L_{Edd}$, and in this case its $L_{X,nuc}$
corresponds to its strongest flare (Baganoff et al. 2001, converted to 
the 0.3--10 keV band).}
\end{figure}

\clearpage

\begin{deluxetable}{lccccccccc}
\tabletypesize{\scriptsize}
\tablecaption{The sample }
\tablewidth{0pt} 
\tablehead{
\colhead{Galaxy} &
\colhead{D} &
\colhead{$M_{BH}$} &
\colhead{Ref.}&
\colhead{$kT$} &
\colhead{$\rho$}&
\colhead{log $L_{X,nuc}$\tablenotemark{a}}&
\colhead{Ref.}&
\\
\colhead{} &
\colhead{(Mpc)} &
\colhead{($10^8$\msun)}&
\colhead{}&
\colhead{(keV)}&
\colhead{($10^{-24}$g cm$^{-3}$})&
\colhead{(erg s$^{-1}$)}&
\colhead{}
\\
\colhead{(1)}&
\colhead{(2)}&
\colhead{(3)}&
\colhead{(4)}&
\colhead{(5)}&
\colhead{(6)}&
\colhead{(7)}&
\colhead{(8)}
}
\startdata
NGC221 (M32)     &0.81 & 0.025$\pm 0.005$   & 1& 0.37$^{+0.28}_{-0.19}$ & 0.13 &   36.44     & 1 \\
NGC821  &24.1 & 0.37$^{+0.17}_{-0.15}$    & 2 & 0.46$^{+0.33}_{-0.25}$  &   0.01$^{+0.027}_{-0.004}$ &   $<$38.66 & 2 \\
NGC1291 &8.9\tablenotemark{b}& 1.1 & 3 & 0.34 &   0.56 &   39.60    & 3 \\
NGC1316 &21.5 & 3.9     & 3 & 0.62$\pm 0.02$ &   0.44 &   39.87    & 4 \\ 
NGC1399 &20.0 & 12     & 3 & 0.8  &   0.47 &   $<$39.14 & 5 \\
NGC1553 &18.5 & 1.6     & 3 & 0.51$^{+0.07}_{-0.08}$ &   0.06 &   40.01    & 6 \\
NGC4261 &31.6 & 5.4$\pm 1.1$     & 4 & 0.6$\pm 0.02$  &   0.17$\pm 0.01$ &   41.15    & 7 \\
NGC4438 &16.1 & 0.5     & 5 & 0.58$^{+0.04}_{-0.10}$ &   0.99 &   39.65    & 8 \\
NGC4472 &16.3 & 7.9     & 3 & 0.8  &   0.32 &   38.69    & 5,9\tablenotemark{c}\\
NGC4486 (M87) &16.1 & 34$\pm 10$  & 6 & 0.8$\pm 0.01$  &   0.36$\pm 0.006$ &   40.88    & 10\\
NGC4594 (Sombrero)& 9.8 & 10$^{+10}_{-7}$  & 7 & 0.65$^{+0.05}_{-0.35}$ &   0.29$\pm 0.10$ &   40.34    & 11\\
NGC4636 &14.7 & 3.0     & 3 & 0.6  &   0.11 &   $<$38.41 & 5\\
NGC4649 &16.8 & 20$^{+5}_{-10}$    & 2 & 0.86$\pm 0.02$ &   1.05$\pm 0.1$ &   38.11    & 12\\
NGC4697 &11.7 & 1.7$^{+0.2}_{-0.1}$     & 2 & 0.33$^{+0.06}_{-0.04}$  &   0.05$\pm 0.01$ &   38.64    & 13\\ 
NGC5128 (Cen A)&4.2 & 2.4$^{+3.6}_{-1.7}$ & 8 & 0.50$\pm 0.05$  &   0.08$\pm 0.01$ &   42.11    & 14\\
IC1459  &29.2 & 25$^{+5}_{-4}$    & 9 & 0.5$\pm 0.1$  &   0.54 &   41.18    & 15\\
IC4296  &49\tablenotemark{d} & 11. & 3 & 0.56$\pm 0.03$ &   1.0$\pm 0.17$  &   41.38    & 16\\
NGC660  &11.8\tablenotemark{e} & 0.22    & 3 & --   &    --  &   38.52    & 17 \\
NGC720  &27.7 & 3.0     & 3 & --   &    --  &   39.19    & 18\\
NGC1332 &22.9 & 10     & 3 & --   &    --  &  $<$38.98  & 19 \\
NGC1407 &28.8 & 4.6    & 3 & --   &    --  &   39.66    & 20 \\
NGC1600 &60\tablenotemark{e}& 12.   & 3 & --   &    --  &  $<$39.83  & 21 \\
NGC2787 &7.48 & 1.2    & 3 & --   &    --  &   38.56    & 22\\
NGC2841 &12.0\tablenotemark{e} & 2.4    & 3 & --   &    --  &   38.52    & 23 \\
NGC3169 &19.7\tablenotemark{e} & 0.7    & 3 & --   &    --  &   41.68    & 22 \\
NGC3226 &23.6 & 1.6    & 3 & --   &    --  &   41.18    & 22 \\
NGC3245 &20.9 & 2.13$^{+1.0}_{-0.6}$    & 10& --   &    --  &   39.03    & 17 \\
NGC3368 &10.4 & 0.14    & 3 & --   &    --  &   39.58    & 24 \\
NGC3489 &12.1 & 0.30    & 3 & --   &    --  &   38.49    & 23 \\
NGC3623 &10.8\tablenotemark{e} & 0.70    & 3 & --   &    --  &   38.86    & 24 \\
NGC3627 & 6.6\tablenotemark{e} & 0.97    & 3 & --   &    --  &  $<$38.08  & 17 \\
NGC4125 &23.9 & 2.4    & 3 & --   &    --  &   39.13    & 24 \\
NGC4143 &15.9 & 4.5    & 3 & --   &    --  &   40.20    & 22 \\
NGC4278 &16.1 & 3.3    & 3 & --   &    --  &   40.61    & 22 \\
NGC4314 &12.8\tablenotemark{e} & 0.16    & 3 & --   &    --  &   38.41    & 24 \\
NGC4321 &16.1\tablenotemark{e} & 0.11    & 3 & --   &    --  &  $<$38.85  & 23 \\
NGC4365 &20.4 & 3.9     & 3 & --   &    --  &  $<$38.18  & 25 \\
NGC4374 &18.4 & 16$^{+20}_{-6}$     & 11& --   &    --  &   39.67    & 26 \\
NGC4382 &18.5 & 1.1     & 3 & --   &    --  &  $<$37.95  & 25 \\
NGC4494 &17.1 & 0.52    & 3 & --   &    --  &   39.12    & 23 \\
NGC4548 &19.2 & 0.28    & 3 & --   &    --  &   40.13    & 22 \\
NGC4552 &15.4 & 4.1    & 3 & --   &    --  &   39.55    & 17 \\
NGC4569 &16.8\tablenotemark{e} & 0.12    & 3 & --   &    --  &   39.67    & 23 \\
NGC4696 &35.5 & 3.2    & 3 & --   &    --  &   40.28    & 24 \\
NGC4826 & 7.5 & 0.55    & 3 & --   &    --  &  $<$38.11  & 23 \\
NGC5846 &24.9 & 3.4    & 3 & --   &    --  &   38.63    & 17 \\
NGC5866 &15.3 & 0.77    & 3 & --   &    --  &  $<$38.59  & 22 \\
NGC6500 &39.7\tablenotemark{e} & 0.67    & 3 & --   &    --  &   40.45    & 22 \\
NGC7331 &13.1 & 0.47    & 3 & --   &    --  &   38.51    & 22 \\
Milky Way&0.008&0.034$\pm0.005$   &12 & 1.3  &    52. &   33.38    & 27 \\
\enddata

\tablenotetext{a}{Nuclear X-ray luminosities refer to the 0.3--10 keV
band; if they were derived in a different band by the
authors in column 8, they have been converted to 0.3--10 keV by using
the spectral shape adopted by these authors.}

\medskip
\tablenotetext{b}{The adopted distance is that of the X-ray analysis paper.}

\medskip
\tablenotetext{c}{$kT$ and $\rho$ have been estimated by 
Ref. 5, while $L_{X,nuc}$ of Ref. 9 has been adopted.}

\medskip
\tablenotetext{d}{This distance has been derived with the SBF method by Mei et al. 2000.}

\medskip
\tablenotetext{e}{The adopted distance refers to $H_0=75$ km s$^{-1}$ 
Mpc$^{-1}$.}

\medskip
\tablerefs{for column 4: (1) Verolme et al. 2002;
(2) Gebhardt et al. 2003; (3) $M_{BH}$ derives from the
$M_{BH}-\sigma$ relation of Tremaine et al. 2002, with $\sigma$ from
McElroy 1995, except for NGC1553, for which $\sigma =186$ km s$^{-1}$ (Longo
et al. 1994); (4) Ferrarese et al. 1996; (5) Machacek et al. 2004;
(6) Macchetto et al. 1997; (7) Kormendy et al. 1996; (8) Marconi et al. 2001;
(9) Cappellari et al. 2002; (10) Barth et al. 2001; (11) Bower et al. 1998;
(12) Sch\"odel et al. 2003.}

\medskip
\tablerefs{for column 8: (1) Ho et al. 2003; (2)
Fabbiano et al. 2004; (3) Irwin et al. 2002;
(4) Kim \& Fabbiano 2003; (5) Loewenstein et
al. 2001; (6) Blanton et al. 2001; (7) Gliozzi et al. 2003;
(8) Machacek et al. 2004; (9) Biller et al. 2004; (10) Di Matteo et
al. 2003; (11) Pellegrini et al. 2003a; (12) Soldatenkov et al. 2003; 
(13) Soria et al. (2005) and Sarazin et al. 2001; (14) Evans et al. 2004 and Kraft et al. (2003); 
(15) Fabbiano et al. 2003; (16) Pellegrini et al. 2003b; (17) Filho et al. 
2004;
(18) Jeltema et al. 2003; (19) Humphrey \& Buote 2004; (20) Zhang \& Xu 2004;
(21) Sivakoff et al. 2004; (22) Terashima \& Wilson 2003; (23) Ho et al. 2001;
(24) Satyapal et al. 2004; (25) Sivakoff et al. 2003; (26) Finoguenov \&
Jones 2001; (27) Baganoff et al. (2003; the X-ray luminosity refers to the 
quiescent state, for the 2--10 keV band).}

\end{deluxetable}

\end{document}